\documentstyle[prl,aps]{revtex}

\begin{document}

\wideabs{

\title{The extraction of $g_A$ from finite volume systems: The long and short
of it }

\author{Thomas D.~Cohen}

\address{Department of Physics, University of Maryland, College Park 20742\\
UMPP\#02-025 \hspace*{.35in} DOE/ER/40762-247 } 

\maketitle

\begin{abstract} Due to a pion-pole contribution, the nucleon axial current
matrix elements can be visualized in position space as becoming delocalized as
the chiral limit is approached, with one-third of the current at distance of
order $1/m_\pi$.  However, this delocalization effect will not cause 
calculations of $g_A$ in a finite system with standard boundary conditions
(periodic for boson fields, antiperiodic for fermions) to have large finite
volume effects.  This is seen by calculating $g_A$ using a nonstandard current
with axial quantum numbers.  The matrix elements of this new current lacks a pion
pole and is not delocalized; however, its zero-momentum matrix element is
identical to that of the standard axial current both for infinite and finite
volumes.  
\end{abstract} }

\medskip

The coupling of the nucleon to axial currents at zero momentum transfer is fixed
by the coupling constant $g_A$.  Due to chiral symmetry and the key role played
by pions in the low-lying dynamics of the nucleon, understanding how $g_A$
emerges from the underlying dynamics of QCD is essential if one wishes to
understand how the low energy dynamics of the nucleon is connected to QCD. 
Attempts to calculate $g_A$ in lattice QCD have given values which are well below
the experimental value\cite{lat1}.  It has recently been argued that this failure
may be due to very large finite volume effects\cite{lat2}.  One may ask whether
there is a basis for understanding the emergence of large finite volume effects
from the structure of QCD.  This issue was brought into sharp focus in a recent
paper by Jaffe\cite{Bob}.  Jaffe made an elegant argument on very general grounds
that the spatial distribution of the axial current in the nucleon is necessarily
spread over large distance as the chiral limit is approached, a phenomenon he
denotes by ``delocalization''.  In particular he showed that as the quark mass
goes to zero, one-third of the axial current is pushed off to infinite distance. 
For finite quark mass, this implies that of order one-third of the axial current
can be found at a distance of order $m_\pi^{-1}$.  It was suggested in Ref.
\cite{Bob} that this effect could lead to large finite volume effects. 
Intuitively it is not implausible that this delocalization of the axial current
could lead to large finite volume effects in calculations of $g_A$: if the pion
is light enough so that large amounts of the axial current are pushed to regions
which are beyond the lattice volume one might expect that the lattice calculation
will miss these contributions leading to substantial errors.

The purpose of this paper is to observe that although ref \cite{Bob} is correct
that one-third of the axial current is delocalized in the chiral limit, the
intuition that this implies large finite volume effects is not (provided the
standard boundary conditions of periodicity for bosons and anti-periodicity for
fermions are imposed).  To make this statement a bit more precise it is necessary
to look a bit more carefully at the heuristic argument why one might expect large
finite volume effects.  The basic issue is the play off between the infinite
volume and chiral limits.  The most serious worry is that as the chiral limit is
approached, one-third of the axial current will be lost outside the box for any
finite sized box.  Thus to get convergence of $g_A$ one needs to choose a box
size significantly larger than one with a length scale set by $1/m_\pi$. 
Mathematically this would  be reflected by  nonuniform limits of the
calculated value of $g_A$ as one approaches the infinite spatial volume and
chiral limits.  Define $g_A^{\small \rm Vol}$ to be the matrix element
of the axial current between nucleon states as calculated with fixed quark mass
(which fixes $m_\pi$) in a finite spatial volume.  The principal concern is
 that $g_A^{\small \rm Vol}$ calculated in the limit
 $V \rightarrow \infty, m_\pi \rightarrow 0$   will only be two thirds
of $g_A^{\small \rm Vol}$ calculated in the limit
$ m_\pi \rightarrow 0, V \rightarrow \infty$.  
However as will be shown below, if the standard boundary conditions are imposed
the limit should be uniform:
 \begin{equation} \lim_{V \rightarrow \infty, m_\pi
\rightarrow 0} \, \, g_A^{\small \rm Vol} \, = \, \lim_{ m_\pi
\rightarrow 0, V \rightarrow \infty,} \, \, g_A^{\small \rm Vol} \label{uniform} 
\; .\end{equation} 
Thus, the major
intuitive concern about large finite volume effects can be put to rest. 
There is a secondary worry, however,  that if much of the strength
of the underlying integrals come from regions comparable to the box size, the
dynamical effects induced by the boundary could alter this part of the
distribution even if the strength is not actually lost. This could lead to
finite volume effects which are larger than those for 
typical baryon matrix elements. 
This secondary concern has not been stated in precise form and, hence, it is hard
to precisely contradict.  However, it will also be shown below there is no reason
based on the delocalization of current discussed in Ref. ~\cite{Bob} to expect
that finite volume effects to be larger for $g_A$ than those for typical matrix
elements.

Let us begin with a brief discussion of the demonstration that the axial current
is delocalized.  In a certain sense, the argument for the delocalization is quite
old;  the result is  well known in the context of mean field chiral soliton
and bag models of the nucleon \cite{models}. In these models the nucleon is
treated as a localized source and the pion field is treated classically. The
localization of the source and the mean field treatments are typically justified
from large $N_c$ QCD arguments. The axial coupling constant is calculated by
integrating the axial current over the nucleon and then matching to the general
Lorentz covariant expression for axial matrix elements assuming that the nucleon
source is localized.  The result is that as $m_\pi \rightarrow 0$  the
integrals which give $g_A$ receives one-third of its contribution from distances
of order $m_\pi$, and as one approaches the chiral limit one-third of the
integrated strength is pushed off to infinity.  In a sense, the crux of the
recent argument of Ref. \cite{Bob} is that these well-known results in the
context of mean field chiral models can be translated into a general
model-independent result with essentially no change in structure.  

To begin, consider the general structure of axial matrix elements of the nucleon
in terms of form factors:
\begin{eqnarray}
\langle&N &(\vec{p'},s',I'_3)|{A}^\mu_a(0)|N(
\vec{p},s,I_3)\rangle = \\ \nonumber
&\overline u&(\vec{p'},s',I'_3)  \frac{\tau_a}{2}
 \left( g_{A}(q^{2})\gamma^\mu \gamma_{5} + h_{A}(q^{2})
q^\mu\gamma_{5} \right ) u(\vec{p},s,I_3)\;, \label{me} 
\end{eqnarray}
 where $u(\vec{p},s, I_3)$ is a free isospin one-half Dirac spinor with a third
component of spin, $s$, and a third component of isospin, $I_3$\cite{axial}.  The
form factors, $g_A(q^2)$ and $h_A(q^2)$, are generally referred to as axial and
induced pseudoscalar form factors, respectively. The axial coupling constant,
$g_A$, is defined as the axial form factor at zero momentum transfer.  The key
point is that the axial current acting on the vacuum can create a pion which
implies that  the induced pseudoscalar form factor has a pion pole
contribution:
 \begin{eqnarray} 
h_A(q^2) & = & h_A^{\small pole}(q^2) + h_A^{\small
short}(q^2) \nonumber \\ h_A^{\small pole}(q^2) & = & - 2 f_\pi g_{\pi \small N N}\;,
\label{ha}
 \end{eqnarray} 
where $f_\pi$ is the pion decay constant (with the
convention that $f_\pi \approx 93 {\rm MeV}$), $g_{\pi \small N N}$ is the
pion-nucleon coupling constant, and $h_A^{\small short}(q^2)$ the non-pion-pole part
of the induced pseudoscalar form factor.  This non-pion-pole contribution to
$h_A$ is denoted as ``short'' to indicate the scale over which it varies
substantially is the hadronic scale, $\Lambda \sim 1 {\rm GeV}$, which is much
larger than $m_\pi$ and hence corresponds to a spatial extent which is much
smaller than $1/m_\pi$.  The pion-pole contribution to the induced pseudoscalar
coupling, however, varies substantially if $q^2$ changes on a scale comparable to
$m_\pi$ which implies contributions on a spatial extent of order $1/m_\pi$.  This
rapid variation of the pion pole term is the origin of the effective delocalization
of the axial current.  

Rather than recapitulate the demonstration of Ref. \cite{Bob} in detail, an
equivalent and somewhat simpler derivation of the same result will be presented
here.  Consider a spatial component of the axial current between spinors.  (A
spatial component is used since the temporal component vanishes at zero momentum
transfer).  At finite momentum transfer the results are frame dependent.  It is
natural to work in the Breit frame of ($\vec p_{\small initial} = -\vec p_{\small
final}$). The Breit frame has the virtue of having momentum transfer but no
energy transfer.  Thus the Fourier transform of the matrix element in the Breit
frame may be directly interpreted in terms of position space; the relative time
variable plays no role.  Introducing the symbol $ G(\vec{q}) $ as the matrix
element of the third spin and isospin components of the axial current between
spin-up protons in the Breit frame and a three-momentum transfer of $\vec{Q}$,
and using Eq.~(\ref{me}) and standard properties of Dirac spinors one has:
\begin{eqnarray}
& G(Q^2)  \equiv 
\langle N(\vec{Q},1/2,1/2)|{A}^3_3(0)|N(-\vec{Q},1/2,1/2)
\rangle = & \nonumber \\  \nonumber \\
 &  \left (  g_A(-|\vec{Q}|^2) \, \left(1+\frac{|\vec{Q}|^2
-2 Q_3^2}{16 M^2}\right ) + h_A(-|\vec{Q}|^2) \frac{Q_3^2}{2 M} \right ) &
 \nonumber \\  \nonumber \\  &\times \frac{ \sqrt{Q^2 + 4 M^2} + 2 M}{4 M}\;,&
\label{G}
\end{eqnarray} 
where $M$ is the nucleon mass.  The choice of the third spin and isospin components 
and the proton spin up states was arbitrary and done for the purposes of
specificity; other choices of components and spin and isospin projections of the
state lead to identical results.  

From Eq.~(\ref{G}) it is immediately apparent that $\lim_{\vec{Q}\rightarrow 0}
G(\vec{Q})  = g_A$, providing that there is no singularity in the induced
pseudoscalar form factor at $q^2=0$.  For any finite $m_\pi$ this is true and the
limit both exists and is uniform in the sense that one can take the $\vec{Q}
\rightarrow 0$ limit from any direction and will get the same answer.  Because
the limit $\vec{Q} \rightarrow 0$ limit is uniform, if we define an angular
average of $G(\vec{Q})$,
 \begin{equation} \overline{G}(Q^2) \equiv \frac{1}{4
\pi} \int {\rm d} \Omega G(\vec{Q}) \;,\label{Gbar}
\end{equation} 
then it is immediately clear that for any nonzero $m_\pi$,
 \begin{equation}
 \lim_{Q
\rightarrow 0} \overline{G}(Q^2)  = g_A \; . 
 \end{equation}

We wish to study $\overline{G}(Q^2)$ for both $q$ and $m_\pi$ well below the
typical hadronic scale of $\Lambda \sim 1\; {\small GeV}$.  Assuming $g_A(q^2)$ and
$h_A(q^2)$ both vary over scales of order $\Lambda$, one sees from 
Eqs. (2),
(\ref{ha}), (\ref{G}) and (\ref{Gbar}) that
 \begin{equation}
\overline{G}(Q^2)  = g_A - \frac{f_\pi g_{\pi NN} }{3 M} \frac{Q^2}{Q^2 +
m_\pi^2} + {\cal O}(Q^2/\Lambda^2, m_\pi^2/\Lambda^2) \; .
 \end{equation} 
Imposing conservation of axial charge in the chiral limit on Eq.~(\ref{me})
 yields the Goldberger-Treiman relation $g_A M = f_\pi g_{\pi N N} + { \cal
O}(m_\pi^2/\Lambda^2)$ \cite{GT} so that 
\begin{equation} 
\overline{G}(q^2)  =
g_A \left (1 - \frac{1}{3} \frac{q^2}{q^2 + m_\pi^2}  \right )+ {\cal
O}(q^2/\Lambda^2, m_\pi^2/\Lambda^2) \; .  \label{del}
\end{equation}
 Note that the preceding equation is the leading order term in chiral expansion;
while there are correction terms of order  $Q^2/\Lambda^2$ and
$m_\pi^2/\Lambda^2$,  $Q^2/m_\pi^2$ has been included to all orders.

The delocalization of axial charge discussed in Ref.~\cite{Bob} is clear from
Eq.~(\ref{del}).  The term proportional to
$Q^2/(Q^2 + m_\pi^2)$ clearly contributes 1/3 of the $Q=0$ result.  It is
also clear that this term varies rapidly over momentum scales corresponding to
spatial variations over long distance scales $\sim 1/m_\pi$.  As $m_\pi
\rightarrow 0$ the spatial distance over which this term contributes becomes
arbitrarily large, {\it i.e.} the current becomes delocalized.  To understand
this as a spatial delocalization of the current in the chiral limit, one needs a
method to visualize the current in position space. The natural way is to define
a spatial axial current density $\rho_A(r)$ as the Fourier transform of
$\overline{G}(q^2)$.  If the current is delocalized in the chiral limit, then
all of the positive moments of $\rho_A(r)$ will diverge.   These moments are
given by 
\begin{eqnarray} \langle r^{2 n} \rangle_A & = &\frac{1}{4 \pi} \int {\rm
d}^3 r \, r^{2 n} \rho_A (r) \nonumber \\
&=&  \prod_{j = 1 , n} (2 + 4 j) (-1)^n \frac{d^n
\overline{G}(Q^2)}{d (Q^2)^n} \; .
 \end{eqnarray} 
Using Eq.~(\ref{del}) one has
\begin{eqnarray}
 \langle r^{2 n} \rangle_A \, & = &\, \frac{g_A \, n! \prod_{j=1,n} (2
+ 4 j)}{3 m_\pi^{2 n}} \, \left ( 1 + {\cal O}(m_\pi^2/\lambda^2) \right )
\nonumber \\ & {\rm for} & \; n
\ge 1 \; .
\end{eqnarray}
all of which diverges $\sim m_\pi^{-2 n}$ as the limit $m_\pi \rightarrow 0$ is approached.
This indicates a delocalized current in 
the chiral limit.

Thus, one-third of the axial current density can be visualized as being
delocalized as the chiral limit is approached.  While this represents an
important insight into the physics, it is important to recall that the
underlying dynamics depends on nothing beyond the fact that the axial current
matrix element has a pion pole contribution.  In this paper, the central issue
is the relationship between the delocalization of the axial current and effects
due to finite spatial volumes. Of course, real lattice calculations have a
number of sources of errors including statistical and discretization errors as
well as errors due to finite spatial and temporal extent. Here, the focus is on
finite spatial volume effects.  To isolate these effects it is useful to
consider the behavior of continuum QCD on a finite three-dimensional volume with
infinite temporal extent.  From the prospect of lattice QCD this can be thought
of as working with a sufficiently fine mesh, a sufficiently large number of
lattice points in the time direction and a sufficiently large number of
configurations so that discretization errors---errors associated with finite
temporal extent---and statistical errors are all negligible.  

The calculation of $g_A$ in finite spatial volume QCD is straightforward in
principle.  We consider the three-dimensional space to be a cube of length $L$,
and impose the boundary conditions that fermion fields are anti-periodic and
boson fields are periodic.  Since the time direction is infinite, one can define
time evolution as being generated by a hermitian hamiltonian operator whose
eigenstates have a well-defined energy.  These energy eigenstates are most
easily treated as being box-normalized.  Because these states are associated
with fields which have well-defined properties under translation in the $x$,
$y$, or $z$ direction by $L$, the energy states are all eigenstates of the
momentum operator with $\vec{p} = 2 \pi L^{-1} (n_x \hat{x} + n_y \hat{y} + n_z
\hat{z})$ (integer $n$'s) for bosons and with $\vec{p} =  \pi L^{-1} ((2 n_x + 1)
\hat{x} + (2 n_y +1) \hat{y} + (2 n_z + 1) \hat{z}$ (integer $n$'s) for fermions.
The box breaks rotational symmetry and hence angular momentum is no longer a
good quantum number.  However, the spin degree of freedom is still well defined
for fermions in the sense that for every momentum there are two degenerate
eigenstates.  Thus the nucleon states can be labeled as $|\vec{p}, I_3, s
\rangle$ with the normalization $\langle \vec{p'}, {I'}_3,s'| \vec{p}, I_3, s
\rangle = \delta_{{I'}_3,I_3} \delta_{s',s} \delta_{{n'}_x,n_x}
\delta_{{n'}_y,n_y}\delta_{{n'}_z,n_z}$. 
One can define the finite volume version of $g_A$ as the
diagonal matrix element of the third space and isospace components of axial
current between spin up proton states on minimum momentum appropriately
normalized to the volume of the box:
 \begin{eqnarray}
 g_A^{\small \rm Vol} &=&   L^3
\langle \vec{p}, {I}_3,s| A^3_3 (\vec{x},t)   |\vec{p}, I_3, s \rangle \nonumber \\
& = & \langle \vec{p}, {I}_3,s|\int_{\small  \rm box}  {\rm d}^3 x \, A^3_3
(\vec{x},t)|\vec{p}, I_3, s \rangle \;, \label{finite}
 \end{eqnarray}
 where the second form of Eq.~(\ref{finite}) stems from translational invariance
of the matrix element with these boundary conditions.

Before proceeding it is useful to distinguish between the method
of calculating $g_A$ for the finite spatial system discussed above as an 
integrated quantity
in a state of good energy from the method actually used in lattice calculations.
In lattice calculations one never calculates exact states of good energy.
 Nucleon matrix elements such as $g_A$ are calculated
as the ratio of a three-point correlation function to a two-point function with
the times of the three currents well seperated in time.  As the seperation in time
between these currents becomes large, the system becomes dominated exponentially by
the lowest energy state with quantum numbers which overlap those of the initial and 
final currents.  As the time seperation between all three currents
goes to infinity the ratio of the three-point function to the two-point function
asymptotically approaches the matrix element in the nucleon state considered above. 
Thus the matrix derived using the states of good energy are equivalent to  
the ratio of the
three-point functions to the two-point functions used in lattice calculations 
proven that
the times are arbitrarily long.  If the times are not arbitraily long there are finite
corrections.

The issue of concern is whether the delocalization of the axial current for the
infinite volume theory leads to large finite volume errors.  It is easy to show
that this is not the case by considering the matrix element of a new axial
current operator.  Suppose that we can find a current operator which in the
infinite volume limit does not suffer from the delocalization effect seen in the
conventional axial current.  If it can be shown that the integrated matrix
element of this new current in the finite volume theory precisely equals
$g_A^{\small \rm Vol}$, then one can replace the original finite volume
calculation of $g_A$ with one based on the new current.  Doing this has the
virtue that delocalization does not occur in the new calculation and hence
cannot be responsible for large finite volume effects.  Since $g_A^{\small \rm
Vol}$ equals the finite volume QCD matrix element of the new current integrated
over space, it follows that $g_A^{\small \rm Vol}$ also cannot have a large finite
volume effect due to delocalization.  It is straightforward to find a current
that has these properties. Since the delocalization is traceable to the pion
pole, we need to find a current whose nucleon matrix elements lack such a
pole.  

Consider the current ${A'}^\mu_a$ defined by
 \begin{equation}
{A'}^\mu_a \equiv \left ( 1  + \frac{\partial^\alpha \partial_\alpha}{m_\pi^2} \right )
 {A}^\mu_a \;, \label{new} 
\end{equation}
where $\partial^\alpha \partial_\alpha$ is the D'Alembertian operator. 
Although this current has no well-defined chiral limit, it is well defined for
any finite value of the quark mass and this is sufficient for the present
purpose.  The matrix elements of this current can easily be expressed in terms
of the old one:
 \begin{eqnarray}
&N(\vec{p'},s',I'_3)|{A'}^\mu_a(0)|N( \vec{p}s,I_3)\rangle  = & \nonumber \\
& \overline
u(\vec{p'},s',I'_3)\frac{1}{2}\langle \tau_a\left({g'}_{A}(q^{2})\gamma^\mu
\gamma_{5} + {h'}_{A}(q^{2}) q^\mu\gamma_{5}\right)u(\vec{p},s,I_3) & \nonumber \\
&{g'}_A(q^2)  =  \left (1 - \frac{q^2}{m_\pi^2}\right ) g_A(q^2) &\nonumber\\
&{h'}_A(q^2)  =  \left (1 - \frac{q^2}{m_\pi^2}\right ) h_A(q^2) \; .& \label{mep}
\end{eqnarray}
It is clear from this form that as $q \rightarrow 0$, ${g'}_A\rightarrow g_A$ 
and ${h'}_A \rightarrow 0$. For the infinite volume case, one can define the function
$\overline{G'}(Q^2)$ in direct analogy to $\overline{G}(Q^2)$,
\begin{eqnarray}
\overline{G'}(Q^2) &\equiv&
 \frac{1}{4 \pi} \int {\rm d} \Omega \,  G'(\vec{Q}) \nonumber \\ 
G'(\vec{Q}) &\equiv&
\langle N(\vec{Q},1/2,1/2)|{A'}^3_3(0)|N(-\vec{Q},1/2,1/2)\rangle \; .  
\end{eqnarray} 
Inserting the matrix elements for $A'$ from
Eq.~(\ref{mep}) into the definition for $\overline{G'}(Q^2)$ gives
$\overline{G'}(q^2) = (1 + {Q^2}/{m_\pi^2} ) \overline{G}(Q^2) $.  Since,
Eq.~(\ref{del}) gives $\overline{G}(q^2)$ to lowest chiral order ({\it i.e.} up
to corrections of order $m_\pi^2/\Lambda^2$ or $Q^2/\Lambda^2$ but including
$Q^2/m_\pi^2$ to all orders) we immediately have the lowest order expression for
$\overline{G'}(Q^2)$:
 \begin{equation}
 \overline{G'}(Q^2) = g_A \left ( 1 +
\frac{2 Q^2}{3 m_\pi^2} \right ) + {\cal O}(Q^2/\Lambda^2, m_\pi^2/\Lambda^2) \;
.  \label{GP}
\end{equation} 
>From the form of Eq.~(\ref{GP}) it is immediately
clear that for any finite quark mass, $\overline{G'}(0) = \overline{G}(0) = g_A$.
Thus, in principle one can use either ${A'}_\mu^a$  or $A_\mu^a$ to extract
$g_A$.

The form of Eq.~(\ref{GP}) shows that the current associated with ${A'}_\mu^a$
is {\it not} delocalized in the sense that a finite fraction of the integrated
strength is at distance scales of order $1/m_\pi$.  Were it so delocalized, all
positive moments of the distribution would diverge as $m_\pi \rightarrow 0$ with
$\langle r^{2 n} \rangle_{A'} \sim 1/m_\pi^{2 n}$ .  These moments are given by
\begin{equation}
 \langle r^{2 n} \rangle_{A'} = = \prod_{j = 1 , n} (2 + 4 j)
(-1)^n \frac{d^n \overline{G'}(Q^2)}{d (Q^2)^n} \; , 
\end{equation}
 which from Eq.~(\ref{GP}) implies that
 \begin{eqnarray} \langle r^{2 }
\rangle_{A'} & = & - \frac{4}{m_\pi^2} \,  ( 1 + m_\pi^2/\Lambda^2 ) \nonumber \\
\langle r^{2 n} \rangle_{A'} &<& {\cal O}(m_\pi^{-2n}) \; \; {\rm for} \;  n > 1 \; \; . 
\label{loc} \end{eqnarray}
Thus none of the higher moments diverge with $m_\pi$ in a fashion consistent
with a delocalized distribution.

One might ask how the $r^2$ moment could diverge with $1/m_\pi^2$ if the
distribution is localized.  The answer is simple: the spatial distribution
$\rho_{A'}(r)$ could be of the following form: 
\begin{eqnarray}
\rho_{A'}(r) &=& g_A \Lambda^3 f_1(\Lambda r) \, +  \,\frac{\Lambda^6}{m_\pi^4} \, \partial_r f_2(\Lambda^2 r/m_\pi ) \label{rho'}\ \\
&{\rm with} & \; \; \int {\rm d}r \, 4 \pi\,  r^2 f_1(r) = 1 \nonumber
\end{eqnarray}
where $f_2(x)$ goes to zero as $x \rightarrow \infty$ faster than any power
law. The first term in Eq.~(\ref{rho'}) integrates to the full value of $g_A$
while the second term integrates to zero due to the derivative.  Moreover, the
second term, although of large overall scale, is of very short range.  Thus it
leads to the moments of Eq.~(\ref{loc}).  It is clear that as the chiral
limit is approached the second term remains sizable only over very short
distances and cannot be associated with delocalization.  

On the other hand, it is straightforward to see that if one uses ${A'}_\mu^a$ to
calculate $g_A$ on a finite volume one gets exactly the same result as
if one uses $A_\mu^a$.  To see this consider
\begin{eqnarray}
&g_{A'}^{\small \rm Vol} \equiv
\langle \vec{p'}, {I'}_3,s'|\int_{\small \rm box} \! \! 
 {\rm d}^3 x {A'}^3_3 (\vec{x},t)|\vec{p}, I_3, s \rangle 
\nonumber \\ 
&=\langle \vec{p'},
 {I'}_3,s'|\int_{\small \rm box}\! \! {\rm d}^3 x  \, \left ( 1  +
 \frac{\partial^\mu \partial_\mu}{m_\pi^2} \right ) 
{A}^3_3 (\vec{x}, t))|\vec{p}, I_3, s \rangle & \; .
\label{finitep}
 \end{eqnarray}
The second equality shows that the matrix element of ${A'}_3^3$ has two
parts---the matrix element of $A_3^3$ and a total derivative term.  This second
term can be written as 
\begin{eqnarray}
&\langle& \vec{p'}, {I'}_3,s'|\int_{\small \rm box} \! \! \! \! \! {\rm d}^3 x  
\frac{\frac{\partial^2}{\partial t^2} - \nabla^2}{m_\pi^2} {A}^3_3 (\vec{x},
t=0))|\vec{p}, I_3, s \rangle  \\
 &= &\frac{L^3}{m_\pi^2} \langle \vec{p'},
{I'}_3,s'|\int_{ \small \rm surface}
 \! \! \! \! \!  \! \! \!{\rm d}^2 x  \,  \hat{n} \cdot \vec{\nabla}
A^3_3(\vec{x},t=0) |\vec{p}, I_3, s \rangle \;,  \nonumber
\end{eqnarray}
where $\hat{n}$ is an outward oriented normal vector;
 the time derivative vanishes since our matrix element is evaluated for an
energy eigenstate and the divergence theorem is used to replace the volume
integral by a surface integral.  On the other hand, the standard boundary
conditions implies that $A_3^3(\vec{x},t) = A_3^3(\vec{x} + L \hat{x},t) =
A_3^3(\vec{x} + L \hat{y}) = A_3^3(\vec{x} + L \hat{y}) $ which in turn means
that $\nabla A^3_3(\vec{x},t=0)$ on one wall of the box is identical to $\nabla
A^3_3(\vec{x},t=0)$ on the opposite wall.  Since the outward normal is opposite
on opposite walls the integral vanishes.  Only the first term of
Eq.~(\ref{finitep}) contributes and thus one has $g_{A'}^{\small \rm Vol} =
g_{A}^{\small \rm Vol}$.

This completes the demonstration.  The current ${A'}_\mu^a$ unlike ${A}_\mu^a$
remains localized in the chiral limit and hence cannot suffer from large finite
volume effects due to delocalization.  However, it gives the same result for
$g_A$ on a finite system as the original current provided the standard boundary
conditions are imposed.  Thus $g_A$ calculated on a finite volume from
${A}_\mu^a$ also cannot have large finite volume effects due to delocalization. 
The proceeding demonstration shows that the intuition that the delocalization of
the axial current implies large finite volume effects is not correct.

It is useful to understand in a more heuristic manner why this intuition fails. 
Let us recall that the underlying reason for concluding the current was
delocalized in the chiral limit was the rapid variation of $G(Q^2)$ for small
pion masses assoicated with the pion pole.  To see what is happening, it is
sensible to restate the intuition that large finite volume effects might be
expected in terms of momentum space.  If one visualizes restricting the nucleon
to a box, one might think that the states will not have well-defined momenta and
the characteristic momenta in the state will be of order $1/L$.  In such a case
the matrix element would have components of momentum with $Q \sim 1/L$ and the
rapid variation of $G(Q^2)$  might be expected to yield large finite volume
effects unless $m_\pi L >> 1$.    However, this intuition is faulty.  In
particular, the perception that the momentum is not well-defined, while valid
for boundary conditions in which the fields vanish on the walls, is {\it not}
valid with the standard boundary conditions (periodic for bosons; antiperiodic
for fermions).  As discussed above, however, the standard boundary conditions
are compatible with states of good momentum.  The effect of imposing the
standard boundary conditions does not mix momenta.  Instead these boundary
conditions allow momentum eigenstates but restrict the allowable momenta. 
Thus, with these boundary conditions one can have $\vec{Q}=0$ exactly, and the
fact that matrix elements vary rapidly with $\vec{Q}$ {\it near} $\vec{Q}=0$ 
becomes irrelevant.  The heuristic argument why delocalization in the chiral
limit does not lead to large finite volume effects depends critically on the use
of the standard boundary conditions. This is consistent with the more formal
demonstration above in which standard boundary conditions were also essential in
showing that the surface integral vanishes.

It is also useful to understand the failure of the intuition that for finite
volumes and a delocalized source one expects to lose some of the strength out of
the box and hence to have large finite volume effects.  Again the boundary
conditions provide the explanation.  The key point is that with these boundary
conditions, a nucleon in the box is equivalent to an infinite system with a
cubic array  (with a lattice spacing of $L$ ) of boxes  each containing a single
nucleon.  Thus, while strength can be lost from a box into its neighbors due to
delocalization, a compensating strength may be gained from strength escaping
into the box from its neighbors.

In summary, it has been shown that the delocalization of the axial current
discussed in Ref.~\cite{Bob} does not lead to large finite volume effects.  This
was shown explicitly by constructing a current which is manifestly not
delocalized and which leads to the same finite volume value for $g_A$ as the
standard current.  The demonstration depended on the imposition of the standard
boundary conditions.  It was argued at a heuristic level why it is not
surprising that delocalization does not yield large finite volume effects
provided these boundary conditions are imposed.

This result leaves open the question of why present lattice calculations
substantially underpredict $g_A$.  Recent lattice results do suggest large
finite volume effects \cite{lat2}.  This does not contradict the present result.
Here it is only shown that the delocalization effect will not lead to large
finite volume corrections. Of course, other effects can cause big finite volume
effects.  Given the relatively small volumes used for the present generation of
lattice calculations it is not too surprising that large finite volume effects
are present.  However, there is also no reason to connect them with the
delocalization effect of 
Ref.~\cite{Bob}.  Ultimately, reliable lattice
calculations should become available on larger lattices and the playoff of the
chiral and infinite volume limits can be explored in the calculation of $g_A$. 
>From the arguments presented here, one expects such calculations to indicate
that the combined limit is uniform.

TDC gratefully acknowledges discussions with J-W Chen, X. Ji and R.~L Jaffe.  
This work was supported in part by the U.S.~Department of Energy.


\end{document}